\documentclass[epj,nopacs]{svjour}
\usepackage{latexsym}
\usepackage{graphics}
\usepackage{slashed}
\begin{document}
\title{Graphene membranes and the Dirac-Born-Infeld action}
\author{James Babington
\thanks{\emph{email:} james.babington@grenoble.cnrs.fr}%
}                     
%
%
\institute{Laboratoire de Physique et Mod\'{e}lisation des Milieux Condens\'{e}s, \\
Universit\'{e} Joseph Fourier and CNRS, \\
Maison des Magist\`eres, \\
38042 Grenoble,\\
France.}
\date{Received: date / Revised version: date}
%
\abstract{
We propose the use of the Dirac-Born-Infeld action in the phenomenological description of graphene sheet dynamics and interactions. Both the electronic properties of the Dirac fermions and the overall dynamics can be incorporated into this model. Classical static configurations, as well as quantum fluctuations of the membrane degrees of freedom can be studied in this framework. This makes it an interesting tool for Casimir physics and novel QED processes.
} 
\maketitle

\section{Introduction}
The Dirac-Born-Infeld (DBI) action~\cite{Born:1934gh,Dirac:1962iy}, originally written down as a way of extending classical electrodynamics and addressing the electron's self energy, is an essential tool if one wants to consider the dynamics of higher dimensional objects e.g. strings and membranes. Its use in string theory has been especially important for the case of Dirichlet-branes (D-branes)~\cite{Polchinski:1995mt,Johnson:2003gi}. These are non-perturbative extended objects which have the ends of fundamental open strings ending on them. At the same time, the low energy dynamics of the brane are described simply by a gauge theory living on this surface. There are two interesting features we wish to draw attention to here. Firstly, the degrees of freedom of the brane couple naturally to the fields in the bulk. Secondly, the microscopic theory can be used to calculate perturbatively the coupling constants between the brane and bulk fields. A natural question arises as to whether this type of dynamics can be applied to other systems. 

Graphene has the interesting property that the degrees of freedom describing the low energy electronic properties are relativistic Dirac fermions~\cite{geim2007,katnelson2007,geim2009,Neto2009}. This together with its one-atom thickness allow in principal for novel applications. The Dirac action allows one to calculate physical observables such as the electrical conductivity or the density of states. If we now want to consider the larger dynamics of the graphene sheet, not just the electronic properties, we need to be able to describe the graphene membrane as a whole~\cite{NetoMembrane08}. Macroscopic graphene sheets are available~\cite{graphenemembrane2008,largegraphen2010} that are in principle dynamical objects in their own right. This together with the tension and torsion~\cite{deJuan2010625,graphenetension2008} necessitates the need for a suitable dynamical theory that captures the relevant field theory aspects in a unified fashion.

In this paper we consider using the DBI action (and suitable generalisations thereof) as a phenomenological tool for studying the low energy dynamics of graphene sheets. In particular, the low energy sector should not be able to resolve the hexagonal lattice structure of graphene (for perturbative field theory calculations the lattice structure will act as an ultraviolet cutoff), whereby collective coordinates can simply be written down. One novel and new feature for this situation is the presence of real fermionic degrees of freedom - the quasi Dirac particles require the DBI action to be modified for their inclusion. This may be seen as a completion of the DBI action (which usually involves only bosonic fields) that has physical relevance and can be performed in a manner outlined in~\cite{Gibbons:2001gy}. The low energy description of the hexagonal lattice (i.e. its position in space) trades the lattice structure and the individual nuclei simply for a world sheet coordinate and an embedding. In a similar fashion, the quasi Dirac particles become the fermionic completion of this embedding. One final factor that can be included is the coupling to the bulk electromagnetic field. This is incorporated as a polarisation type term that lives on the membrane itself, as well as from a covariant derivative of the fermions. We shall then work out some of the consequences  due to the nature of their mutual couplings.

The outline of this paper is as follows. In Section~\ref{sec:basics} we describe the basics of the DBI action to fix notations and conventions. In Section~\ref{sec:graphene} we generalise this appropriately to include the Dirac fermions on the membrane and possible couplings between the membrane itself and the bulk electromagnetic field. In Section~\ref{sec:bosonic} we perform an evaluation of the DBI action for a dielectric background that has only a one dimensional dependence and a simple graphene drum placed in a constant magnetic field. In Section~\ref{sec:fermionic} an expansion of the fermionic part of the action is performed to obtain a modified Dirac equation due to the presence of the polarisation term. This paves the way for three calculations - the spectrum of Landau levels, an induced Chern-Simons term and finally a modified Schwinger effect. In Section~\ref{sec:casimir} a path integral picture is presented whereby a static potential between between two parallel graphene membranes is obtained by considering the S-matrix. Finally in Section~\ref{sec:conclusion} we draw our conclusions.

\section{Basics of the DBI action}
\label{sec:basics}
\begin{figure}[htbp]
\begin{center}
\resizebox{0.3\textwidth}{!}{
\includegraphics{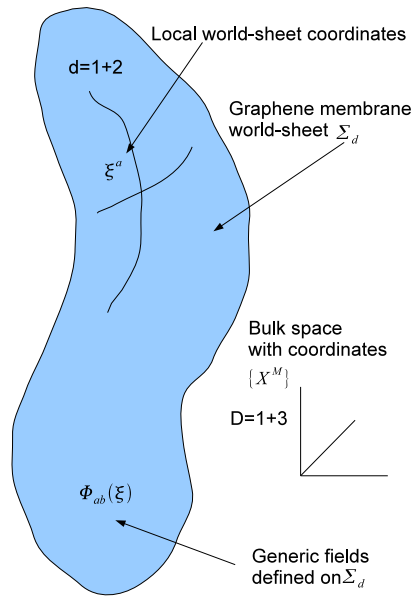}
}
\caption{The $d=1+2$ membrane embedded in the $D=1+3$ bulk space. The coordinates $\xi^a$ are the (in general) curved world sheet indices of the membrane, whilst the $X^M$ are the curved bulk coordinates. The membrane is then given as the embedding $X^M(\xi)$. Generic fields $\Phi_{ab}(\xi)$ can be defined on the membrane.}\label{fig:membrane}
\end{center}
\end{figure}

To fix our notation and conventions, we briefly describe what the DBI action consists of in the relativistic setting of field theory. Consider a $d=1+2$ membrane (two spatial dimensions, one time) embedded in the ambient $D=1+3$ bulk spacetime (three  spatial dimensions, one time, see Figure~\ref{fig:membrane}). The DBI action is then essentially the area of this space (denoted by the manifold $\Sigma_3$). One such model takes the form
\begin{equation}
\label{eq:DBIaction}
S_3=T_3\int_{\Sigma_3}d^3\xi \sqrt{\vert \det [g_{ab} + \mathcal{F}_{ab}]\vert }.
\end{equation}
The quantities and conventions used here are: $T_3$ is the tension (or mass per unit area) of the membrane; $\xi^{a}$ are the curved world sheet coordinates of the membrane with index $a=0,1,2$; the embedding coordinates of the membrane are $X^M(\xi^a)$ in the bulk space with the index $M=0,1,2,3$; and a bulk metric $G_{MN}(X)$ field~\cite{Johnson:2003gi}. With this structure the bulk metric is said to be pulled back to the world sheet with the relation $g_{ab}=G_{MN}\partial_a X^M\partial_bX^N$ and we can now calculate the area of the membrane. We have also included a $d$-dimensional $U(1)$ gauge field on the membrane given by $\mathcal{A}_a$, leading to the field strength $\mathcal{F}_{ab}=(d\mathcal{A})_{ab}$ (which is invariant under a gauge transformation). This is a simple illustration of how one can also include into the action fields that live on the membrane. We will do this in particular for the Dirac fermions that are to describe the electronic properties of the membrane as well as for the electromagnetic field strength itself after performing an appropriate pull-back.

Using the reparametrisation invariance of the DBI action, one can choose a gauge (sometimes called the static gauge or the Monge gauge) where the world-sheet coordinates coincide with the physical coordinates of the membrane leaving one coordinate orthogonal to the membrane. This describes the embedding of the membrane in the bulk space:-
\begin{eqnarray}
\label{eq:mongegauge}
X^{0}(\xi)&=&\xi^{0},\; X^{1}(\xi)=\xi^{1}, \nonumber \\
 X^{2}(\xi)&=&\xi^{2},\; X^{3}=Z(X^0,X^1,X^2).
\end{eqnarray}
With this choice of gauge one sees that the DBI action takes the form of a non-linear field theory on the membrane. There is one scalar field that describes the embedding of the membrane in the bulk space. One can consider fluctuations around this embedding by expanding the determinant and the square root for a particular background metric. The expansion then gives a fully interacting field theory as a derivative expansion which can then be quantised. It also affords a description of minimum energy surfaces for static configurations of membranes that are subjected to some potential via the bulk metric.

\section{The graphene DBI action}
\label{sec:graphene}

We now seek to write down a phenomenological version of the DBI action for a graphene membrane. The degrees of freedom of the membrane are coupled to a set of bulk fields providing the necessary interactions. This is naturally a low energy description where the hexagonal lattice structure of graphene is simply replaced with the embedding coordinates described above, just as the fermions are replaced with the Dirac spinor fields. The underlying lattice structure also provides an UV cutoff for any of the effective field theory correlation functions that are calculated on the membrane. 

Indeed in~\cite{Bordag:2009fz}, the field equation for the bulk photon is written down with the addition of a static defect term to account for the interaction with the graphene fermions; it takes the form
\begin{equation}
\partial_{M} F^{MN}+\delta (X^{3}-z_0)\Pi^{MN}A_{M}=0.
\end{equation}
Here $\Pi^{MN}$ is the one-loop polarisation operator for the photon on the graphene sheet extended to $D=(1+3)$ dimensions and $z_0$ is the position of the graphene sheet in the $X^3$ direction. As it stands the graphene sheet is a static object that interacts only via its fermions. Our task is now to make the whole object dynamical by constructing a suitable low energy DBI action.

The degrees of freedom on the membrane are the membrane coordinates $X^{M}(\xi^a)$ themselves, together with the $d=1+2$ spinors $\psi(\xi^a)$ that describe the fermion-hole pairs on the graphene sheet itself. Both of these interact with the bulk $D=1+3$ electromagnetic potential $A_M$. The first quantity to form is a spinorial covariant derivative on the membrane
\begin{equation}
D_a \psi=\partial_a \psi-e A_M\partial_aX^{M} \psi-\frac{1}{4}\omega^{\;bc}_a[\Gamma_{b},\Gamma_{c}] \psi, \label{eq:covariantderiv}
\end{equation}
where $\omega_{a}$ is a one-form spin connection (compatible with the pull-back metric $g_{ab}$), and $\Gamma^a$ are the $d$ dimension gamma matrices that are rescaled with the Fermi velocity $v_F$ as in~\cite{Bordag:2009fz}. The inclusion of torsion terms as discussed in~\cite{deJuan2010625} is also possible should it need to be present. The membrane phonons, the bulk photon field and the curvature of the membrane can be incorporated in this way. In graphene there are four species of fermions, so we should really include an extra index on the spinor to account for this; however we will omit this and reinstate the four species content when it is appropriate.

A further coupling of the photon field to the membrane coordinates (with coupling constant $\lambda$) in the form of the pull-back of the bulk field strength is also implemented. It resembles the $\mathcal{F}_{ab}$ term included in Equation~(\ref{eq:DBIaction}) and represents the surface area polarisation energy~\cite{Wang20093050,PhysRevB.75.045407}. The simplest action one can write down for a graphene membrane interacting with a bulk metric and electromagnetic field is then
\begin{eqnarray}
\label{eq:DBIaction2}
 S_G=\int_{\Sigma_3}d^3\xi (\vert \det [g_{ab} +i\bar{\psi}\Gamma_{a} (D_{b}+\frac{m}{3}\Gamma_{b})\psi  
 +\lambda F_{ab}] \vert )^{1/2}, \nonumber \\
\end{eqnarray}
where $F_{ab}=F_{MN}\partial_{a}X^{M}\partial_{b}X^{N}$ is the pull-back of the bulk electromagnetic field strength. We have also allowed for a small  fermion mass $m$ in the above (the factor of three is included for the correct normalisation of the fermion mass in the resulting Dirac equation). One should note as well that we have two light cone structures; the light cone for the electronic properties is given by the Fermi velocity, whilst for a bulk vacuum we retain the normal Minkowski light cone. As it stands we have not taken into account that the membrane is an elastic object. The elastic properties can be included by replacing Equation~(\ref{eq:mongegauge}) with 
\begin{equation}
X^M = \delta_{\; a}^{M}\xi^a + T_{\; N}^M\Phi^N(\xi), \label{eq:fluctcoords}
\end{equation}
where the mixed tensor $T_{\; N}^M$ contains the  elastic stress tensor parallel and normal to the membrane and the $\Phi^N(\xi)$ are the fluctuations of the embedding coordinates (the phonon modes). In this way one can include the acoustic modes (longitudinal and transverse) and their propagation in a full manner.

It is also a simple matter to include a Chern-Simons term into the action using the same pull-back principles. Indeed with a view to including such interactions as in~\cite{Fialkovsky:2009wm,PhysRevD.81.065005}, one can write down (for some constant $\rho$ to be determined)
\begin{eqnarray}
S_{GCS}&=&\rho\int_{\Sigma_3}A_a F_{ab}\epsilon^{abc} \nonumber \\
&\equiv &\rho\int_{\Sigma_3}d^3\xi A_M F_{NP}\partial_a X^M \partial_b X^N \partial_c X^P \epsilon^{abc}
\end{eqnarray}
which is then added to the dynamical DBI graphene action. In fact as we will see later, we can evaluate this coupling constant as a one loop radiatively induced term upon integrating out the fermions.

The coupling constant $\lambda$, as mentioned previously, has the interpretation in the full action as a static surface polarisability due to the effective dipole moments that arise from different carbon bonds~\cite{Wang20093050,PhysRevB.75.045407}.  By performing an expansion in Equation~(\ref{eq:DBIaction2}) to quadratic order in the field strengths for static weak fields, one has the schematic relation $\lambda^2\sim \alpha_{surface}$. In contrast, the vacuum polarisation $\Pi^{MN}$ in~\cite{Bordag:2009fz} results from integrating out the fermions. The significance of this coupling is also the way by which the photons can interact with the phonon on the graphene sheet when considering fluctuating fields. Indeed, from Equation~(\ref{eq:DBIaction2}) we see that it is possible to generate three and four point vertices that involve both the photon and the phonon modes in a self consistent way, similar to what is found in~\cite{basko2009}. The main focus for the rest of the paper will be on how the $\lambda F_{ab}$ term couples to the other fields and derive their implications in a few cases of practical interest.

\section{Applications and phenomenology of the purely bosonic sector: classical aspects}
\label{sec:bosonic}

We can distinguish two cases of practical interest in using this action. The first is to perform a derivative expansion of the action where the membrane degrees of freedom are split into background plus fluctuations. The second is where one finds a solution of the full field equations for a given set of known bulk fields. In both cases we need to be able to calculate the determinant of the pull-back fields. In this section we will consider only the bosonic membrane coordinates as the degrees of freedom where we set $\langle \psi \rangle = 0$. Firstly we will consider a toy model of magneto-electric media in the bulk to gain some familiarity with the pull-back of fields. After this we consider fluctuations of the membrane with imposed boundary conditions and the effect of the $\lambda$ term.

\subsection{Magneto-electric media in the bulk and a static configuration}
\label{sec:media}

An interesting situation to consider is when we have the bulk fields propagating in a background magneto-electric media, with an underlying flat metric. For the case when the permittivity $\epsilon_{ij}(t,x)$ and permeability $\mu_{ij}(t,x)$ are local functions of spacetime (purely spatial indices $i,j=1,2,3$), we will replace the bulk metric $G_{MN}$ with both electric and magnetic tensors in a simple multiplicative way. This naturally produces a local light cone structure in the bulk and should properly be thought of as a toy model, since we are neglecting all dispersion related properties and we putting them in by hand. The pull-back of the metric $g_{ab}$ in this scheme we choose to be
\begin{eqnarray}
g_{ab}= -\partial_{a}X^{0}\partial_{b}X^{0}+(\epsilon \mu)^{-1}_{ij}\partial_{a}X^{i}\partial_{b}X^{j}.
\end{eqnarray}

Consider now a static configuration of a membrane immersed in some dielectric with $A_M=0$. The gauge choice of world sheet coordinates are $X^0=\xi^0$, $X^1=\xi^1$,  $X^2=\xi^3$, and $X^3= Z(\xi)$. In addition, we take the permeability to be $\mu_{ij}=\delta_{ij}$ and the permittivity to be $\epsilon_{ij}=\mathrm{diag}(1,1,\epsilon(Z))$, so that there is only a profile in the $X^3$ direction.
With this choice the pulled back metric becomes
\begin{eqnarray}
g_{ab}=\nonumber \\
\begin{scriptsize} 
  \left(\begin{array}{ccc}
-1+\epsilon^{-1}(Z)\partial_t Z\partial_t Z & \partial_1Z \partial_t Z &\partial_2Z \partial_t Z  \\ 
\partial_1Z \partial_t Z&  1+\epsilon^{-1}(Z)\partial_1 Z\partial_1 Z& \partial_1Z \partial_2 Z\\
\partial_2Z \partial_t Z& \partial_1Z \partial_2 Z & 1+\epsilon^{-1}(Z)\partial_2 Z\partial_2Z 
\end{array} 
\right).\end{scriptsize}  \nonumber \\
\end{eqnarray}
One can now easily calculate the determinant (and square root) whereby the action thus becomes
\begin{eqnarray}
S_G&=&\int d^3\xi[ -1+ \epsilon^{-1}(Z)[(\partial_t Z)^2- (\partial_1 Z)^2+(\partial_2 Z)^2] \nonumber \\
&&+\mathcal{O}(Z^4)]^{1/2}.
\end{eqnarray}
The general scheme of things should then be clear. The scalar field will have non-trivial dynamics due to background media (in this case dielectric) and the higher derivative terms. Since it is a static configuration $\partial_t Z=0$. Suppose it takes the form of an infinite strip of graphene of width $l$ that lies in the $(X^1,X^3)$ plane. The equation of motion simplifies when the action is constrained, with the area taking some constant value (implying the constancy of the effective Lagrangian).  Let the model of the permittivity function profile be given by $\epsilon(Z)=e^{Z/L}$ to be definite. Then the equations of motion simplify to
\begin{equation}
\partial_{1}e^{-Z/2L}=c_1.
\end{equation}
 This can be simply integrated to
\begin{equation}
Z=-2L\ln (c_1X^1+c_2).
\end{equation}
What one can draw from this is that the precise shape of graphene membranes will be dictated by the nature of the static background fields the graphene couples to. As alluded to beforehand, this is not a realistic model. Nevertheless, it is interesting to see that a surface profile can be calculated once a background is known. More complicated solutions (Catenoids) of this nature can be found in~\cite{Gibbons:1997xz}.

\subsection{A graphene drum-skin in an external magnetic field}
\label{subsec:drumskin}

The fact that the bulk electromagnetic field couples to the membrane in a non-trivial fashion implies that we should be able to see its effect in simple phenomenology. To this end we calculate the shift of resonant frequencies for a graphene sheet that has a disk (drum-skin) configuration. This is subjected to a constant magnetic field in a direction normal to the graphene disk.

Consider again the expansion of the action in the static gauge detailed subsection~\ref{sec:media} to quadratic order in the fields for only the normal direction (that is the $X^3=Z$ direction) fluctuations together with $F_{12}=B_3=B\neq 0$. We will also include the speed of sound $v$ of the normal modes:-
\begin{eqnarray}
 \det (g_{ab}+\lambda F_{ab})&=&[-1+v^2\dot{Z}^2-(\partial_1 Z)^2-(\partial_2 Z)^2] \nonumber \\
& +& [-1+v^2\dot{Z}^2]\lambda^2(F_{12})^2.
\end{eqnarray}
This leads to an equation of motion for the transverse oscillations
\begin{equation}
v^2(1+\lambda^2B^2)\ddot{Z}-\nabla^2 Z=0.
\end{equation}
One recognizes now a standard drum skin problem. In polar coordinates the solution takes the form $Z(t,r,\theta)=e^{i\omega t} X(r,\theta)$ where
\begin{eqnarray}
X(r,\theta)&=&\sum_{n=0}a_nJ_{n}(kr)e^{in\theta} +c.c., \\
k^2&=&v^2\omega^2(1+\lambda^2B^2).
\end{eqnarray}
For a drum skin of radius $R$, we have the boundary condition $X(R,\theta)=0$, whereby the frequencies now take on a discrete form given by the zeroes of the Bessel function $J_n(k_{mn}a)=0$ such that
\begin{eqnarray}
\omega^2_{mn}=\frac{k^2_{mn}}{v^2(1+\lambda^2B^2)}.
\end{eqnarray}
In particular, the lowest resonant frequency is shifted to $\omega_{00}\approx 2.4/(av^2(1+\lambda^2B^2)$.

\section{Fermionic sector applications and phenomenology: quantum aspects}
\label{sec:fermionic}
One can see that the fermionic sector is particularly interesting for the electronic properties of the membranes. We have a system in which to study the properties of quantum field theories in a low-energy and low-dimension window. The DBI action Equation~(\ref{eq:DBIaction2}) can viewed as a 'machine' for producing couplings between the bosonic and fermionic sectors, and amongst themselves. As by way of providing some examples, we shall investigate the effects of the couplings between the fermionic terms and the field strength $\lambda F_{ab}$ in Equation~(\ref{eq:DBIaction2}). In particular, we shall focus attention on the scenario where we have static external electric and magnetic fields present together with fluctuating fermionic fields. The fermions will be considered in both the first and second quantised scenarios.

\subsection{First quantised fermions in background electromagnetic fields: Landau levels}

Before moving on to a specific example let us make a few remarks about the mechanics of the DBI action Equation~(\ref{eq:DBIaction2}) involving spinors. It is possible to consider the fermions on a general curved background (for an overview of how curvature affects the electronic properties see~\cite{Vozmediano10,Vozmediano:2008zz}) and with background electromagnetic fields. The DBI action needs to be expanded to quadratic order in the fermion fields from which the Dirac equation can be found as the leading contribution. To see this, consider the matrix expansion
\begin{eqnarray}
\det (\textbf{M} +\textbf{N})&=& \det (\textbf{M})\det (\textbf{1}+\textbf{M}^{-1}\textbf{N}) \nonumber \\
 &\approx & \det (\textbf{M})(1+\mathrm{Tr}  (\textbf{M}^{-1}\textbf{N})),
\end{eqnarray}
which is true when the absolute values of the matrix elements of $\textbf{N}$ are much smaller than those of $\textbf{M}$. This immediately yields the massless Dirac equation (for $\lambda = 0, m=0$) on a curved space associated with the pulled back membrane metric $g_{ab}$ in a fully dynamical manner
\begin{eqnarray}
\label{eq:diraceqn}
\Gamma ^{a}D_a \psi &=& \Gamma ^{a}(\partial_a \psi-e A_M\partial_aX^{M} \psi 
-\frac{1}{4}\omega^{\;bc}_a \Gamma_{ab}\psi ) \nonumber \\
&=&0.
\end{eqnarray}
Here $\Gamma_{ab}=[\Gamma_{b},\Gamma_{c}]$. It has been used in~\cite{Gonzalez:1992qn,Pincak2005267} to calculate the energy spectrum of fullerene molecules when the curved space is just the two dimensional sphere or a spheroid. From now on we consider only flat geometries where $\omega^{\;bc}_a=0$.

An interesting application of the model is when there are background electromagnetic fields present, arising from the $\lambda$ term. The effect of this is to provide a novel type of coupling. Heuristically speaking, just as the metric couples to the energy momentum tensor of the fluctuating fields, the background electromagnetic field will couple to the antisymmetric part of the fermion pull-back. 

As an example involving a first quantized field (with $m=0$), consider the situation of a plane graphene membrane (in the x-y plane $\mathbf{R}^2$) that is in a constant magnetic field $B$ (in the z-direction normal to the graphene plane). In this case the expansion is performed about the matrix
\begin{equation}
( g_{ab}+\lambda F_{ab})|_{ \mathbf{R}^2}=\left(\begin{array}{ccc}
-1 & 0  & 0 \\ 
0&  1 & \lambda B  \\
0& -\lambda B   & 1
\end{array} 
\right).
\end{equation}
To linear order in $\lambda$ we find a modified Dirac equation
\begin{equation}
(\Gamma^aD_{a}\psi+\lambda B(\Gamma_{1}D_{2}-\Gamma_{2}D_{1})\psi)  = 0.
\end{equation}
For $\lambda = 0$ one finds the usual relativistic Landau energy levels and wave functions~\cite{Neto2009}. The energy levels are given by $E(N)=\pm \hbar (\sqrt{2 eB/c})v_F \sqrt{N}$, where $N=0,1,2,\cdots$, whilst the corresponding spinorial eigen-functions are given in terms of Hermite polynomials. By performing a first order perturbative calculation due to the $\lambda$ term, one finds a zero energy shift for a given energy level. It is therefore necessary to go to at least quadratic order to see its effect in connection with the Landau levels.

\subsection{Fermionic fluctuations in background electromagnetic fields: induced Chern-Simons terms and pair production}

As a further example we will consider the fermionic field as a second quantised field.
Upon looking at the expansion of the DBI action, we see that in addition to the standard gauge coupling term $eA_a\bar{\psi} \Gamma^a \psi$ we have the spin current interaction $(\lambda m/6)F_{ab}\bar{\psi} \Gamma^{ab} \psi$. If we compute the one loop effective action for these two interactions as in~\cite{Dunne:1998qy}, we find a three index polarization tensor $\Pi^{abc}(p)$ given by
\begin{eqnarray}
\Pi^{abc}(p)=\int \frac{d^3k}{(2\pi)^3}\mathrm{tr}\left[\Gamma^{a}\frac{(\slashed p+\slashed k) -m}{(p+k)^2+m^2}\Gamma^{bc}\frac{(\slashed k) -m}{k^2+m^2}\right]. \nonumber \\
\end{eqnarray}
One can extract the low energy behaviour ($p \rightarrow 0$) from this and indeed it is similar to that of the standard  $d=1+2$ induced Chern-Simons term. The effective action reads
\begin{equation}
S_{GCS}[A,m,\lambda] = -i\frac{\lambda e m^3}{3\pi \vert m \vert v^2_F}\int A_{a}F_{bc}\epsilon^{abc}.
\end{equation}
The presence of such terms are of interest when one wishes to consider observables such as polarisation rotation effects~\cite{Fialkovsky:2009wm} and Casimir-Polder forces~\cite{PhysRevD.81.065005}.

As a final example involving the fermion as a quantum field, one can ask how the Schwinger effect~\cite{Schwinger:1951nm} is modified due to the presence of the $\lambda$ coupling. Consider the situation where we have constant externally applied electric field $E$ in the plane of the membrane and the effect this has on the fermions when they are taken to be second quantized fields. The Schwinger effect for this system has been described in~\cite{Allor:2007ei} for a constant electric field, together with a proposed experiment to measure the pair production in terms of macroscopically observed transient currents. The result found there for the rate of production per unit area per unit time of fermion-hole pairs is
\begin{equation}
w^{1+2}=\frac{(eE)^{3/2}}{\pi^2\hbar^{3/2}v^{1/2}_F}\sum^{\infty}_{n=1}\frac{1}{n^{3/2}} \exp \left[ -\frac{n \pi m^2 v_f^3}{eE\hbar} \right].
\end{equation}
The DBI action Equation~(\ref{eq:DBIaction2}) provides us with a new set of couplings. To see the effect of the mass term and the polarization term, we must go back to the effective action~\cite{Lin:1999,Itzykson:1980rh} that is obtained from expanding Equation~(\ref{eq:DBIaction2}) to quadratic order in the fermionic fields
\begin{eqnarray}
\ln W[A]= \nonumber \\
 -\mathrm{Tr} \ln \left(\frac{\Gamma^a P_a - m +i\epsilon}{\Gamma^a( P_a -eA_a) -(\lambda m/6) \Gamma^{ab}F_{ab} - m +i\epsilon} \right).
\end{eqnarray}
One can evaluate this perturbatively as an expansion in $\lambda$. As for the case of Landau levels, there is no contribution at linear order in $\lambda$. The first contribution arises at quadratic order and is found to be
 \begin{equation}
w^{1+2}=\frac{(eE)^{3/2}}{\pi^2\hbar^{3/2}v^{1/2}_F}\sum^{\infty}_{n=1}\frac{1}{n^{3/2}} \exp \left[ -\frac{n \pi m^2(9+\lambda^2 E^2) v_f^3}{9eE\hbar} \right].
\end{equation}
At least for this coupling the effect would be to lessen the pair production for increasing electric field strength because the polarisation term requires work to be done.

\section{Casimir energies }
\label{sec:casimir}

One way of evaluating the DBI action is to perform an expansion in the fluctuating membrane fields around a background. It can be thought of as a 'test particle' or probe both in a classical or quantum picture. In particular, one may be interested in how the membrane reacts when it interacts with other matter. One such scenario is provided by Casimir physics~\cite{Casimir:1948dh} and the subtleties therein. For example, in the case of two  parallel plates with a general dielectric media in between them, there is a possible stress within the dielectric media itself that arises due to a particular choice of stress tensor~\cite{raabe:013814}. Thus one would expect that a graphene membrane placed in such a media as a probe would be subjected to an observable force and could be used to distinguish between different theoretical proposals. See also~\cite{Bordag:2009fz,gomezsantos-2009-80,bitbol2010} for related considerations.

We will follow here a simpler program and confine ourselves to calculating a standard two body Casimir energy in the framework of QFT. Specifically, by the consideration of scattering amplitudes it is possible to calculate a static potential between the bodies. 

Suppose we have two infinite parallel planes of graphene separated by some distance $R$, and we would like to know what the Casimir energy is between them. It can be found by calculating the scattering amplitude for the two - membrane $\rightarrow$ two - membrane process. This is what is done for the Casimir-Polder potential between two polarisable particles based on a phenomenological action~\cite{Feinberg:1970zz,Itzykson:1980rh}. The differences here are that  we are dealing with extended (infinite) bodies together with a different form for the effective vertices. There are in principal different contributions to the static potential; notably the atomic surface polarisability as well as the contribution due to virtual fermion-hole pairs. We shall only consider the contribution that arises due to the $\lambda$ term in Equation~(\ref{eq:DBIaction2}).

The basic scattering amplitude we want to evaluate is for two flat and parallel rigid membranes with initial 4-momenta $k^M_1$ and $k^M_2$, and final momenta $k^{\prime M}_1$ and $k^{\prime M}_2$. We therefore need to evaluate the four-point correlation function $\langle X(1)X(2)X(3)X(4) \rangle$. The correlation function is defined using a two body path integral $\mathcal{Z}_D$ given by
\begin{eqnarray}
\mathcal{Z}_D&=&\int [dA] \exp \left( i\int_{D} F^2/4 \right)\mathcal{W}_d[A],  \\
\mathcal{W}_d[A]&:=&\int\prod^{2}_{I=1} \left[dX^{I}d\bar{\psi}^{I}d\psi^{I}\right]
\exp \left( \sum^{2}_{I=1} iS^{(I)}_G \right).
\end{eqnarray}
In this setup, all of the fields are taken to be full quantum fields. We are therefore interested in obtaining another effective action and thereby vertex, where we know how the field strength $F_{MN}$ couples to the membrane $X^M$. The DBI action Equation~(\ref{eq:DBIaction2}) already contains such a set of terms given by the $\lambda$ term. If we expand out the bosonic part to second order in $\lambda$, together with the gauge choice of coordinates Equation~(\ref{eq:fluctcoords}) (taking the elastic matrix to be $T^N_{M}=\delta_M^N$ for simplicity) one can obtain the necessary couplings
\begin{eqnarray}
S[X,F] &=& \int d^3\xi [\vert\det (g_{ab}+\lambda F_{ab})\vert]^{1/2} \nonumber \\
&=&\int d^3\xi \left[1-\frac{1}{2}\partial_a\Phi^M \partial_a \Phi^M - \frac{\lambda^2}{2}F_{ab}F^{ab}\right],  \\
F_{ab} &=& F_{MN}(\delta_a^M+\partial_a\Phi^M)(\delta_b^N+\partial_b\Phi^N).
\end{eqnarray}
In this way one generates both the standard projection of the the full field strength onto the membrane, a cubic coupling and most importantly a quartic coupling of the form
\begin{eqnarray}
S^{(4)}_{int} \sim \int d^3\xi  \lambda^2F^2(\partial\Phi)^2.  
\end{eqnarray}
With this effective vertex it is a simple matter to calculate the interaction energy between the two graphene membranes. The four-point function becomes
\begin{eqnarray}
 \langle X(1)X(2)X(3)X(4) \rangle &\sim & \lambda^4 \int d^3\xi_1 d^3\xi_2  \langle X(1) \partial X \rangle   \nonumber \\
&&\times \langle X(2) \partial X \rangle
 \langle X(3) \partial X \rangle
 \langle X(4) \partial X \rangle\nonumber \\
&&\times \langle F(X(\xi_1)) F(X(\xi_2)) \rangle \nonumber \\
&&\times\langle F(X(\xi_2)) F(X(\xi_1)) \rangle ,
\end{eqnarray}
where $\xi_1$ and $\xi_2$ are the world-sheet coordinates on membrane one and two respectively. From the known form of the time ordered photon propagator, the correlation function $\langle  FF \rangle $ is given by
\begin{eqnarray}
\langle F(X(\xi_1)) F(X(\xi_2)) \rangle \sim \frac{1}{\left[R^2+|\xi_1-\xi_2|^2\right]^2}.
\end{eqnarray}
The remaining two point function of the scalars just describe the in and out legs. It is possible now to calculate the corresponding S-matrix and the static potential. By doing the time integration and then the integration over all the two dimensional membrane space, one derives the static potential density. This can clearly be seen to behave like $V\sim \lambda^4/R^5$. It is interesting to see a different power law behaviour that arises due to such a phenomenological interaction term.

\section{Conclusions}
\label{sec:conclusion}

The intent of this paper has been to set up a simple phenomenological scheme to address the dynamics of graphene sheets as a whole. This includes both the overall dynamical structure of the graphene lattice approximated as a continuum and their electronic properties as described by the inclusion of world sheet fermions. These can be naturally coupled to the bulk electromagnetic field and different schemes exist with which to evaluate the resulting action. One notable feature is the way in which the fermion spinorial fields live on the graphene membrane and couple to bulk fields which in principle provide a consistent scheme for calculating bulk-membrane quantities, e.g. graphene sheet reflection coefficients in a dielectric. The inclusion of a pull-backed electromagnetic field into the action (the $\lambda$ term) as a way of including the static polarisation of the membrane has consequences when we consider field fluctuations. Upon expansion we find cross couplings that produce additional perturbative terms, that in principle have observable consequences. 

An important and obvious next step is to consider the effect of real materials in the bulk and finite temperature. All the fluctuating physical fields need to be Fourier transformed in time as well as the bulk background fields. It would also be interesting to try and apply the DBI formalism directly to scattering experiments in the low energy sector. The calculation of photon scattering amplitudes from a graphene sheet should give the same information as the reflection coefficients obtained from matching conditions. Pursuing a S-matrix programme would be a worthwhile effort to better understand the associated scattering phenomenology.

\begin{acknowledgement}

\section*{Acknowlegements}

I would like to thank D. Basko for useful discussions and insightful remarks, and  J. Gracey for helpful comments.  I would also like to thank G. Rastelli and S. Scheel for useful early discussions. This work was supported by the ANR contract PHOTONIMPULS ANR-09-BLAN-0088-01.

\end{acknowledgement}

\end{document}